\documentclass{article}
\usepackage{spconf,amsmath,graphicx,hyperref}

\usepackage{cite}
\usepackage{amsmath,amssymb,amsfonts}
\usepackage{algorithmic}
\usepackage{graphicx}
\usepackage{textcomp}
\usepackage{xcolor}
\usepackage{booktabs}

\title{Discriminator-Free Adversarial Domain Adaptation for \\ Robust Speech Classification}
\name{Yacouba Kaloga, Ina Kodrasi}
\address{Signal Processing for Communication, Idiap Research Institute, Martigny, Switzerland}
\def\ninept{\def\baselinestretch{.9}\let\normalsize\small\normalsize}

\begin{document}
\ninept
\maketitle
\begin{abstract}
The performance of speech classification systems such as speech emotion recognition or pathological speech detection can degrade substantially when there is a domain shift between training and test data. This shift can arise due to e.g., differences in recording setup, acoustic environment, speaker, language, or speaking style.
Adversarial domain adaptation addresses this problem by encouraging representations that are less sensitive to domain-specific differences. However, many adversarial approaches rely on training a separate domain discriminator, which can be unstable and difficult to train. In this paper, we propose to use a discriminator-free adversarial learning network (DALN) for unsupervised domain adaptation in speech classification tasks. DALN reuses the task classifier as an implicit discriminator and aligns domains through the nuclear-norm Wasserstein discrepancy on classifier outputs. We extensively evaluate the performance of DALN on intra-lingual, cross-lingual, and noisy adaptations for speech emotion recognition and pathological speech detection.  Extensive experimental results show that DALN generally achieves the best performance across the considered tasks and domain shifts, outperforming discriminator-based domain adaptation approaches.
\end{abstract}

\section{Introduction}
Speech signals contain rich information about the speaker, including their affective state, identity, age, and physical and cognitive health~\cite{schu13}. The ability to extract such information from speech has led to a broad range of automatic speech classification tasks, such as e.g., speaker verification~\cite{aloradi2022}, speech emotion recognition (SER)~\cite{geo24,10095892} or pathological speech detection~\cite{11086511,su15043204}.
Self-supervised speech representations have proved remarkably effective for these tasks, with models such as wav2vec~2.0~\cite{baevski2020wav2vec20frameworkselfsupervised} and HuBERT~\cite{hsu2021hubertselfsupervisedspeechrepresentation} providing representations that outperform hand-crafted descriptors~\cite{9964237}.
Despite these advances, speech classification systems can be highly sensitive to domain shifts between training and test conditions. 
Such shifts can arise from differences in background noise, recording setup, language, or elicitation protocol, and can substantially degrade classification performance~\cite{schu,Amiri2024TestTimeAF}. 
Performance sensitivity to domain shift is a major obstacle to practical deployment, because the conditions encountered at test time are often unknown in advance, while collecting labeled data for every target domain is costly. 

Unsupervised domain adaptation addresses this challenge by aligning source and target distributions using \emph{labeled} source and \emph{unlabeled} target data~\cite{ganin2016domainadversarialtrainingneuralnetworks}. 
Existing unsupervised domain adaptation approaches largely fall into two categories, i.e., non-adversarial and adversarial domain adaptation.
Non-adversarial approaches directly match predefined source and target domain
statistics such as feature covariances~\cite{DBLP:journals/corr/SunS16a}, but do not explicitly account for
distributional differences beyond those selected statistics. 
In contrast, adversarial domain adaptation approaches typically learn representations that make the source and target domains difficult to distinguish using an additional discriminator. 
For example, the domain-adversarial neural network (DANN)~\cite{ganin2016domainadversarialtrainingneuralnetworks} uses a marginal domain discriminator, whereas the conditional domain-adversarial network (CDAN)~\cite{NEURIPS2018_ab88b157} conditions a discriminator on classifier predictions.
These approaches have been successfully applied to speech classification tasks, including cross-corpus and noise-robust SER and pathological speech detection~\cite{latif2022selfsupervisedadversarialdomain, wang21u_interspeech}.
However, domain-adversarial approaches can be difficult to optimize. For example, DANN has exhibited convergence difficulties in SER, with discriminator-based objectives suffering from vanishing gradients once the domains become easily separable~\cite{shen2018wassersteindistanceguidedrepresentation}. 
The performance of adversarial approaches can also be sensitive to the adversarial weight and to the relative capacities of the feature extractor and discriminator~\cite{chen2022reusing}.
These issues are particularly relevant when adapting speech representations, where datasets may be very heterogeneous and the amount of data available may be very limited.

These limitations highlight the need for an adversarial formulation that does not require training a separate discriminator.
Inspired by the discriminator-free adversarial learning network~(DALN) proposed for image classification~\cite{chen2022reusing}, in this paper we adapt this formulation to speech classification tasks. To this end, we use representations extracted from frozen self-supervised speech encoders (e.g., wav2vec 2.0~\cite{baevski2020wav2vec20frameworkselfsupervised} or HuBERT~\cite{hsu2021hubertselfsupervisedspeechrepresentation}) followed by a small trainable multilayer perceptron (MLP).
Domain alignment is driven by a nuclear-norm Wasserstein discrepancy computed from the
classifier outputs, which encourages confident and diverse predictions on the
target domain~\cite{cui2020discriminabilitydiversitybatchnuclearnorm}. Since alignment is performed directly on the classifier outputs, a separate domain discriminator is not required. 
\begin{figure*}
\vspace{-0.9cm}
    \centering    \includegraphics[width=0.85\linewidth]{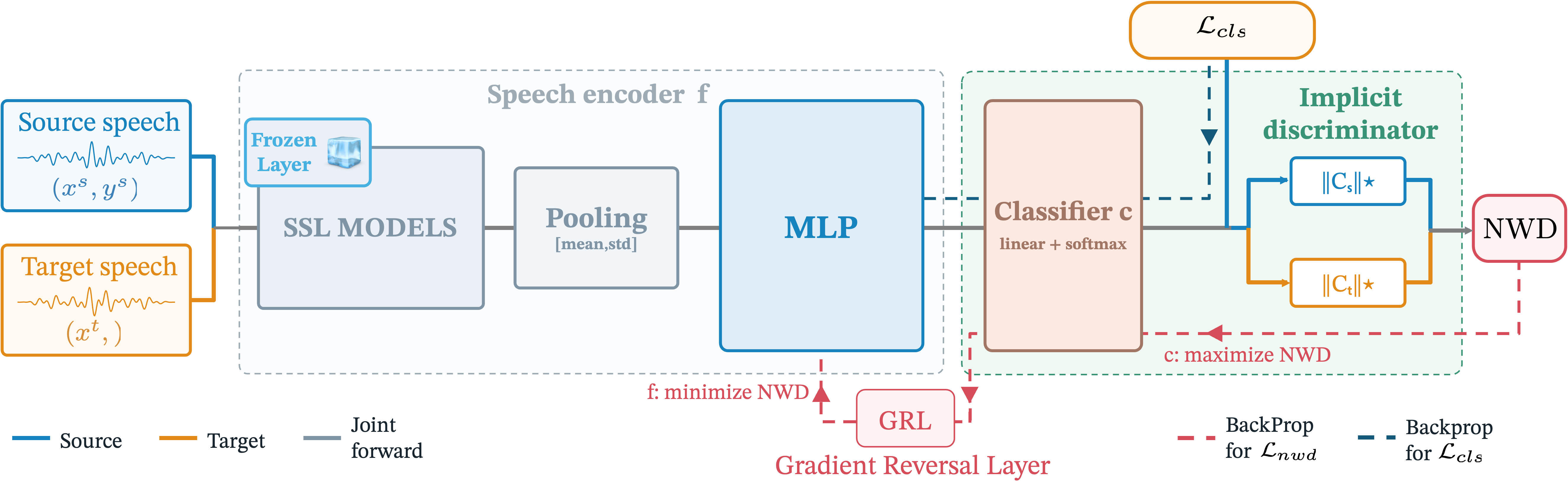}
\caption{\textbf{DALN architecture for robust speech classification.} Source
and target utterances are jointly processed by a frozen SSL model, statistical pooling, and a trainable MLP encoder $f$. The task
classifier $c$ is supervised by the source classification loss
$\mathcal{L}_{\mathrm{cls}}$ and reused as an implicit domain discriminator.
The nuclear-norm Wasserstein loss $\mathcal{L}_{\mathrm{nwd}}$ is computed
from the source and target prediction matrices $C_s$ and $C_t$. Through gradient reversal, the
classifier maximizes this loss while the encoder minimizes it, aligning
the domains without an additional domain discriminator. Solid lines denote the
forward pass and dashed lines the corresponding backward paths.}
    \label{fig:daln_speech}
\end{figure*}

Our contributions are twofold. First, we adapt DALN to variable-length speech inputs using frozen self-supervised representations and a
lightweight trainable adaptation head (cf. Fig.~\ref{fig:daln_speech}).
Second, to our knowledge, we provide the first systematic evaluation of discriminator-free adversarial adaptation across multi-class and binary speech classification tasks such as SER and pathological speech detection. We evaluate the approach extensively on seven emotion recognition datasets under intra-lingual, cross-lingual, and noisy adaptation settings. We further assess its applicability to pathological speech detection under noisy adaptation settings using the PC-GITA dataset of neurotypical and dysarthric speech~\cite{orozco-arroyave-etal-2014-new}.
Across these experiments, DALN generally yields the highest performance among the evaluated adversarial methods, supporting its use as an alternative to DANN and CDAN
for speech classification under domain shift.

\section{Unsupervised domain adaptation}
\label{sec:method}
Let $\mathcal{D}_s = \{(x_i^s, y_i^s)\}_{i=1}^{n_s}$ be a labeled source domain and
$\mathcal{D}_t = \{x_j^t\}_{j=1}^{n_t}$ an unlabeled target domain, drawn from
different distributions over a shared input space $\mathcal{X}$ and label space
$\mathcal{Y} = \{1,\dots,K\}$.
An encoder $f$ maps an input to a representation, and a classifier $c$ maps that representation to class posteriors. The goal of unsupervised domain adaptation is to learn $f$ and $c$ that generalize to $\mathcal{D}_t$, despite the distributional shift between the source and target domains.
\subsection{Domain-adversarial neural network}
\label{sec: dann}
In DANN~\cite{ganin2016domainadversarialtrainingneuralnetworks}, domain adaptation is achieved by introducing a domain discriminator $d$ alongside $f$ and $c$. 
Training combines two losses, i.e., a classification loss on the source domain and an adversarial domain loss.
The classification loss is the standard cross-entropy loss $\ell_{\mathrm{ce}}$ on labeled source data, i.e.,
\begin{equation}
\mathcal{L}_{\mathrm{cls}}(f,c) = \frac{1}{n_s}\sum_{i=1}^{n_s} \ell_{\mathrm{ce}}\bigl(c(f(x_i^s)), y_i^s\bigr),
\label{eq:dann_cls}
\end{equation}
whereas the domain loss is a binary cross-entropy loss over source and target samples, i.e.,
\begin{equation}
\begin{aligned}
\mathcal{L}_{\mathrm{dom}}(f,d) \!\!=\!\!{}&
-\frac{1}{n_s}\sum_{i=1}^{n_s} \log d\bigl(f(x_i^s)\bigr) 
-\frac{1}{n_t}\sum_{j=1}^{n_t} \log\bigl(1 - d(f(x_j^t))\bigr),
\end{aligned}
\label{eq:dann_dom}
\end{equation}
with $d(f(x)) \in (0,1)$ the discriminator's predicted probability that $x$ comes from the source domain.
Training is done through the min-max optimization objective
\begin{equation}
\min_{f,c}  \max_d \left\{\mathcal{L}_{\mathrm{cls}}(f,c) - \lambda \mathcal{L}_{\mathrm{dom}}(f,d) \right\},
\label{eq:dann}
\end{equation}
where $\lambda$ is a weighting parameter that controls the strength of the
adversarial signal. The discriminator $d$ is trained to minimize
$\mathcal{L}_{\mathrm{dom}}$ by better distinguishing source from target
representations, while $f$ is trained to maximize it by producing
representations that fool $d$.
In practice,~\eqref{eq:dann} is not optimized with alternating updates but with a single backward pass using a gradient reversal layer (GRL) inserted between $f$ and $d$. The GRL leaves the forward pass unchanged and reverses the gradient passed back to $f$, so that a single gradient step minimizes $\mathcal{L}_{\mathrm{dom}}(f,d)$ with respect to $d$ and maximizes it with respect to $f$.
Following~\cite{ganin2016domainadversarialtrainingneuralnetworks}, $\lambda$ is not fixed but increased from $0$ to $1$ during training with a sigmoid schedule parametrized by $\alpha$, which aims at suppressing the noisy adversarial signal early in training. Such an adversarial formulation can be sensitive to the relative capacities of $d$ and $f$ as well as to this schedule, potentially resulting in \emph{negative transfer} where adaptation performs worse than the source-only baseline~\cite{ganin2016domainadversarialtrainingneuralnetworks}.

\subsection{Discriminator-free adversarial learning network}
To avoid training a separate discriminator, DALN~\cite{chen2022reusing}, originally proposed for image classification, reuses the task classifier $c$ as an implicit domain discriminator. Its adversarial signal is derived directly from differences between source and target predictions.

Let $x^s$ and $x^t$ denote batches of $B$ source and target inputs, and let
$c(f(x^s)), c(f(x^t)) \in \mathbb{R}^{B\times K}$ denote the corresponding
matrices of classifier softmax predictions. DALN measures the discrepancy
between these matrices using the nuclear-norm Wasserstein discrepancy (NWD),
\begin{equation}
\mathrm{NWD}\bigl(c(f(x^s)),c(f(x^t))\bigr)
=
\Bigl(
\left\|c(f(x^s))\right\|_\star
-
\left\|c(f(x^t))\right\|_\star
\Bigr)/B,
\label{eq:nwd}
\end{equation}
where $\|\cdot\|_\star$ is the nuclear norm, i.e., the sum of singular
values. The nuclear norm of a prediction matrix reflects the confidence and
diversity of its predictions, with large nuclear norm values indicating
confident and diverse predictions and small nuclear norm values indicating
uncertain or collapsed ones~\cite{cui2020discriminabilitydiversitybatchnuclearnorm}.
Hence, a large NWD indicates that the target predictions are less confident
and less discriminative than the source predictions, reflecting a gap between the source and target domains. DALN uses the NWD directly as an adversarial loss, i.e., 
$\mathcal{L}_{\mathrm{nwd}}(f,c) = \mathrm{NWD}\bigl(c(f(x^s)),c(f(x^t))\bigr)$
and trains $f$ and $c$ jointly with the
objective
\begin{equation}
\min_{f,c} \left\{\mathcal{L}_{\mathrm{cls}}(f,c) + \lambda\, \max_c \mathcal{L}_{\mathrm{nwd}}(f,c) \right\},
\label{eq:daln}
\end{equation}
where $\mathcal{L}_{\mathrm{cls}}$ is the source classification loss in~\eqref{eq:dann_cls} and $\lambda$ controls the strength of the adversarial signal. A GRL is inserted between $f$ and $c$ on the
NWD branch, so that a standard gradient step on~\eqref{eq:daln} minimizes
$\mathcal{L}_{\mathrm{nwd}}$ with respect to $f$ while maximizing it with
respect to $c$. Hence, the classifier is encouraged to amplify the distinction between source and target predictions, whereas the encoder is encouraged to learn representations that reduce this distinction. 
\subsection{DALN for speech classification}
\label{sec:daln_speech}

Figure~\ref{fig:daln_speech} summarizes our adaptation of DALN to speech. We
retain its discriminator-free objective from~(\ref{eq:daln}) unchanged, but replace the
end-to-end trainable image backbone in~\cite{chen2022reusing} with a frozen self-supervised speech
encoder, mean-and-standard-deviation temporal pooling, and a small trainable
MLP. Freezing the encoder avoids the cost of fine-tuning a $300$M-parameter
model (such as e.g., wav2vec~2.0 or HuBERT) separately for every target condition,
making adaptation faster and more suitable for practical deployment.
Hence, the encoder $f$ in Eq.~\eqref{eq:daln} comprises the frozen speech model,
temporal pooling, and the trainable MLP, while $c$ is a linear classifier
followed by softmax. 
Consequently, adaptation updates the MLP and classifier, while leaving the self-supervised encoder frozen.

\section{Experimental Settings}
\label{sec:setup}
To evaluate the performance of DALN for unsupervised domain adaptation in speech classification, we use the following experimental settings.

\noindent{\bf Tasks and datasets.} \enspace We evaluate DALN on two exemplary speech classification tasks, i.e.,
multiclass SER and binary pathological speech detection, using the datasets summarized in Table~\ref{tab:corpora}. For SER, we use
seven publicly available datasets spanning a range of recording styles and languages, allowing us to test adaptation across diverse domains. These datasets span eight classes (\emph{anger}, \emph{boredom}, \emph{disgust}, \emph{fear}, \emph{happiness}, \emph{neutral}, \emph{sadness}, and \emph{surprise}), with each dataset covering only a subset of these classes.
For pathological speech detection, we use the PC-GITA dataset of neurotypical speakers and patients with Parkinson's disease~\cite{orozco-arroyave-etal-2014-new}, which is a commonly used
dataset for this task~\cite{11086511}.

\noindent{\bf Domain shifts.} \enspace We consider two sources of
domain shift, i.e., \emph{corpus shift} and \emph{noise shift}.

\emph{Corpus shift} evaluates cross-dataset adaptation under intra-lingual
and cross-lingual settings. For the intra-lingual setting, we consider
three English SER datasets, i.e., \textit{CREMA-D}~\cite{cao14},
\textit{IEMOCAP}~\cite{Busso2008IEMOCAP:Interactiveemotionaldyadic}, and
\textit{MSP-PODCAST}~\cite{mario14}. Each dataset is adapted to each of the other two, yielding
six examples of intra-lingual adaptation. For this analysis, we exclude utterances from emotion classes that are not shared across all three datasets, retaining only anger, happiness, neutral, and sadness. For the cross-lingual setting, we consider four datasets across four languages, i.e., CREMA-D (English), \textit{CaFE}~\cite{gournay18} (French), \textit{EMOVO}~\cite{costantini-etal-2014-emovo} (Italian), and \textit{EmoDB}~\cite{865cd1db96a2481da142ee142e8d01bc} (German). We adapt each dataset to each of the remaining three, yielding 12 cross-lingual adaptation examples. For each example, we retain only utterances from emotion classes shared by the source and target datasets. The shared classes are anger, disgust, fear, happiness, neutral, and sadness, except for the the CaFE-to-EMOVO and EMOVO-to-CaFE adaptations, for which surprise is also shared.

\emph{Noise shift} is evaluated individually for every SER and pathological
speech dataset, each retaining its original class set. The target domain in this case is formed by corrupting the source with additive noise from
the DEMAND database~\cite{thiemann2013diverse}. Each source dataset is adapted to
$20$ different target domains, generated by combining one of four noise types (living room,
meeting room, cafeteria, or street traffic) with one of five SNRs ($0$, $5$, $10$, $15$, or $20$~dB).
\begin{table}[t!]
\centering
\caption{Characteristics of the considered datasets in terms of language (Lang.), speaking style, number of speakers (Spk.), number of utterances (Utt.), and number of classes (Cl.).}
\label{tab:corpora}
\setlength{\tabcolsep}{4pt}
\scalebox{0.85}{\begin{tabular}{lllrrc}
\hline
Corpus & Lang. & Style & Spk. & Utt. & Cl. \\
\hline
CREMA-D~\cite{cao14}
    & EN & acted, scripted & 91 & 7442 & 6 \\
IEMOCAP~\cite{Busso2008IEMOCAP:Interactiveemotionaldyadic}
    & EN & acted, scripted/impro. & 10 & 5531 & 4 \\
MSP-PODCAST~\cite{mario14}
    & EN & natural, podcast & 102 & 12053 & 4 \\
EmoFilm-EN~\cite{6aad4e4b46a34cacaa78e013e73b0a4c}
    & EN & acted, film & 77 & 343 & 5 \\
CaFE~\cite{gournay18}
    & FR & acted, scripted & 12 & 936 & 7 \\
EMOVO~\cite{costantini-etal-2014-emovo}
    & IT & acted, scripted & 6 & 588 & 7 \\
EmoDB~\cite{865cd1db96a2481da142ee142e8d01bc}
    & DE & acted, scripted & 10 & 535 & 7 \\
\hline
PC-GITA~\cite{orozco-arroyave-etal-2014-new}
    & ES & read & 100 & 1100 & 2 \\
\hline
\end{tabular}}
\end{table}

\noindent \textbf{Input representations.} \enspace
We use task-specific self-supervised speech representations for the two classification tasks. For SER, we extract embeddings from HuBERT-large~\cite{hsu2021hubertselfsupervisedspeechrepresentation}, which has been shown to perform well for emotion recognition~\cite{wang2022finetunedwav2vec20hubertbenchmark}. Among its 
$24$ transformer layers, we select layer 
$12$ for embedding extraction, as it achieves the highest mean UAR in our preliminary speaker-independent emotion-recognition experiments on EmoDB (without domain adaptation). For pathological speech detection on PC-GITA, we use embeddings from layer 
$10$ of wav2vec~2.0-large~\cite{baevski2020wav2vec20frameworkselfsupervised}, following~\cite{purohit2025automatic}.

\noindent \textbf{Baselines.} \enspace
We compare DALN with two discriminator-based adversarial approaches, i.e., DANN~\cite{ganin2016domainadversarialtrainingneuralnetworks} and
CDAN~\cite{NEURIPS2018_ab88b157}.
CDAN conditions its discriminator on the classifier predictions and is included here as a stronger and class-aware
extension of DANN.
This allows us to assess whether DALN's performance stems
merely from class-conditioned alignment or specifically from removing the explicit discriminator. All methods share the same architecture for $f$ and $c$.

\noindent \textbf{Evaluation.} \enspace
We follow a cross-fold speaker-independent evaluation protocol. 
For SER datasets, speakers are partitioned into disjoint test groups using an adaptive rule, i.e., each group contains one speaker for datasets with at most six speakers, two speakers for datasets with seven to
12 speakers, and $20\%$ of the total number of speakers (rounded down) for larger datasets. Each group
forms one test fold, and the number of folds used for each dataset is the number required for
every speaker to occur in the test set exactly once. During hyperparameter selection, $20\%$ of the remaining
speakers (with a minimum of one) are held out for validation.
After hyperparameter selection, these speakers are merged back into the training set for final training and evaluation.
For the PC-GITA dataset, we use the standard speaker-independent 10-fold cross-validation~\cite{sheikh2024graphneuralnetworksparkinsons}. Performance is evaluated
using unweighted average recall (UAR) for SER and speaker classification accuracy for pathological speech detection. For each adaptation method, we report performance as the difference in target test set performance between the adapted model and the unadapted baseline model trained on the source domain. Experiments are repeated with three different random seeds, and we report the mean performance difference across these repetitions. Due to space constraints, we report only the standard deviation averaged over seeds and adaptation settings for each method. For each setting, we additionally assess whether an adaptation method's advantage over a competing method is statistically significant using a two-sided paired t-test, treating each source–target pair as one paired observation. Because several source–target pairs share a corpus or set of speakers, and are therefore not fully independent, these tests should be interpreted as exploratory.

\noindent \textbf{Training.} \enspace
Models are trained with AdamW~\cite{loshchilov2019decoupledweightdecayregularization} (weight decay $10^{-4}$, constant learning rate) for $30$ epochs with a batch size of $32$.
All hyperparameters are tuned on \emph{the source  validation set} only, so that target labels are never used at any stage.
Hyperparameter tuning is done for each task and each dataset in two stages. We first select the classification architecture by training on the source domain and evaluating performance on the source validation set, sweeping the learning rate $\in \{10^{-3},10^{-4}\}$, the number of hidden layers $\in \{2,3,4\}$, and the width of hidden layers $\in \{64,128\}$. Keeping the selected architecture fixed, we then tune one adaptation hyperparameter per method, again on the source validation set, i.e., the weight $\lambda \in \{0.1,0.5,1,2\}$ in Eq.~\eqref{eq:daln} for DALN and its equivalent for CDAN, and, for DANN, the parameter $\alpha \in \{1,5,10,20\}$ of the schedule of $\lambda$ in Eq.~\eqref{eq:dann}. For DALN and CDAN, the gradient reversal layer follows the fixed warm-start schedule of the original DALN implementation~\cite{chen2022reusing}.

For cross-corpus adaptations, the adaptation hyperparameters are separately tuned for each source--target pair.
For noisy adaptations, the adaptation hyperparameters are tuned once for each source dataset using cafeteria noise at an SNR of 10~dB, and the resulting values are reused across all 20 noisy target domains.  
Architecture and adaptation hyperparameters are selected based on one repetition and held fixed for the remaining repetitions.

Note that the original DALN approach uses a fixed adaptation parameter $\lambda=1$~\cite{chen2022reusing}. Our preliminary results suggest that this robustness of DALN to the exact value of $\lambda$ carries over to the speech domain.
Nevertheless we tune $\lambda$ in this paper, so that every method has its adaptation hyperparameter tuned in the same way ($\lambda$ for DALN and CDAN, $\alpha$ for DANN).

\section{Experimental Results}
\label{sec: exp_res}

In the following, we present the obtained results for the considered intra-lingual (cf.~Table~\ref{tab:intra}), cross-lingual (cf.~Table~\ref{tab:cross-lingual}), and noisy adaptation (cf.~Tables~\ref{tab:ser_noise} and~\ref{tab:pcgita_noisy}) settings. For each setting, we report the mean difference in target performance between the adapted model and the unadapted baseline, together with the standard deviation across seeds averaged over all adaptations.

\begin{table}[t!]
\centering
\vspace{-0.5cm}
\footnotesize
\caption{Change in target UAR (\%) for the six intra-lingual adaptations using DANN, CDAN, and the proposed approach. The UAR of the baseline unadapted model on the source test set (Source) and on the target test set (Target) is also presented.}
\label{tab:intra}
\setlength{\tabcolsep}{7.5pt}
\scalebox{0.9}{\begin{tabular}{lccccc}
\hline
Adaptation & Source & Target & DANN & CDAN & Proposed \\
\hline
CREMA-D $\to$ IEM   & 86.4 & 50.1 & $+$5.3 & $+$6.8 & \textbf{$+$7.6} \\
CREMA-D $\to$ MSP   & 86.4 & 39.2 & $-$2.3 & $-$0.5 & \textbf{$-$0.3} \\
IEM $\to$ CREMA-D   & 66.4 & 57.6 & $+$0.8 & $+$7.3 & \textbf{$+$16.7} \\
IEM $\to$ MSP       & 66.4 & 38.5 & $-$1.1 & $+$0.3 & \textbf{$+$1.2} \\
MSP $\to$ CREMA-D   & 37.4 & 36.8 & $-$6.5 & $-$8.1 & \textbf{$+$4.0} \\
MSP $\to$ IEM       & 37.4 & 38.0 & $+$3.4 & $+$6.8 & \textbf{$+$14.4} \\
\hline
\emph{Average std}  &      &      & 3.0 & 4.0 & 3.2 \\
\emph{Average}         &      &      & $-$0.1 & $+$2.1 & \textbf{$+$7.3} \\
\hline
\end{tabular}}
\end{table}

\begin{table}[t]
\footnotesize
\centering
\caption{Change in target UAR (\%) for the 12 cross-lingual adaptations using DANN, CDAN, and the proposed approach. The UAR of the baseline unadapted model on the source test set (Source) and on the target test set (Target) is also presented.}
\label{tab:cross-lingual}
\setlength{\tabcolsep}{6.5pt}
\scalebox{0.9}{%
\begin{tabular}{lccccc}
\hline
Direction & Source & Target & DANN & CDAN & Proposed \\
\hline
CaFE $\to$ CREMA-D
  & 73.9 & 29.4 & $+$5.2 & \textbf{$+$22.3} & $+$11.1 \\
CaFE $\to$ EmoDB
  & 72.0 & 51.3 & $-$25.6 & $-$0.1 & \textbf{$+$13.2} \\
CaFE $\to$ EMOVO
  & 72.5 & 34.5 & $-$2.6 & $+$2.8 & \textbf{$+$8.5} \\
CREMA-D $\to$ CaFE
  & 75.2 & 42.6 & $-$12.1 & $-$9.6 & \textbf{$-$3.5} \\
CREMA-D $\to$ EmoDB
  & 75.2 & 58.0 & $-$6.7 & $-$6.9 & \textbf{$+$6.5} \\
CREMA-D $\to$ EMOVO
  & 75.2 & 42.0 & \textbf{$-$1.8} & $-$6.9 & $-$7.9 \\
EmoDB $\to$ CaFE
  & 92.5 & 30.2 & $-$12.1 & \textbf{$-$2.5} & $-$3.1 \\
EmoDB $\to$ CREMA-D
  & 92.5 & 30.9 & $-$10.9 & $-$7.4 & \textbf{$+$2.9} \\
EmoDB $\to$ EMOVO
  & 92.5 & 49.3 & $-$18.2 & $-$9.8 & \textbf{$-$2.4} \\

EMOVO $\to$ CaFE
  & 60.4 & 37.1 & $-$11.3 & $+$4.2 & \textbf{$+$7.8} \\
EMOVO $\to$ CREMA-D
  & 62.8 & 32.1 & $+$12.1 & $+$7.2 & \textbf{$+$14.1} \\
EMOVO $\to$ EmoDB
  & 67.3 & 72.1 & $-$14.4 & $+$4.4 & \textbf{$+$13.2} \\
\hline
  \emph{Average std}
  & &  & 3.4 & 3.8 & 3.6 \\
\emph{Average}
  & &  & $-$8.2 & $-$0.2 & \textbf{$+$5.0} \\
\hline
\end{tabular}}
\end{table}
\textbf{Intra-lingual adaptation for SER.} \enspace
Table~\ref{tab:intra} shows that DALN is the strongest adversarial method in all six intra-lingual adaptations for SER, improving over the unadapted model in five of them. The average performance improvement across all intra-lingual adaptations for DALN is 7.3\%, compared with $2.1$\% for CDAN and $-0.1$\% for DANN. DALN's advantage
over DANN is statistically significant at a significance threshold of $0.05$ ($t(5)=3.03$, $p=0.029$). Its advantage over CDAN is smaller and more variable across adaptations, and the t-test does not indicate statistical significance at this sample size ($t(5)=2.45$, $p=0.058$).
However, DALN outperforms CDAN in all six adaptations, which is unlikely to occur by chance (i.e., a two-sided exact sign test yields $p=0.031$).
Overall, DALN is the most effective and consistent of the three adaptation approaches in all considered intra-lingual adaptations for SER.

\textbf{Cross-lingual adaptation for SER.} \enspace
Similarly to the intra-lingual setting, DALN is generally the strongest adversarial method for cross-lingual SER adaptation, as shown in Table~\ref{tab:cross-lingual}.
DALN achieves an average performance improvement across all cross-lingual adaptations of $5.0$\%,
compared with $-0.2$\% for CDAN and $-8.2$\% for DANN. Treating the 12 cross-lingual adaptations
as paired observations, DALN's advantage is statistically significant over both
DANN ($t(11)=3.93$, $p=0.002$) and CDAN ($t(11)=2.62$, $p=0.024$) at a significance threshold of $0.05$. 
As can be observed, DALN improves the performance over the unadapted model in eight of the 12 cross-lingual SER adaptation examples, compared to only five for CDAN and only two for DANN.

\begin{table}[t!]
\centering
\vspace{-0.5cm}
\footnotesize
\caption{Change in target UAR (\%) for the noisy adaptations on SER datasets using DANN, CDAN, and the proposed approach. The UAR of the baseline unadapted model on the source test set (Source) and on the target test set (Target) is also presented.}
\label{tab:ser_noise}
\setlength{\tabcolsep}{4pt}
\scalebox{1}{\begin{tabular}{lccccc}
\hline
Corpus & Source & Target & DANN & CDAN & Proposed \\
\hline
CaFE          & 72.6 & 37.1 & $+$0.0  & $+$12.3 & \textbf{$+$15.5} \\
CREMA-D       & 75.2 & 54.4 & \textbf{$+$5.4} & $+$1.9 & $+$0.2 \\
EmoDB         & 91.7 & 76.9 & $-$19.6 & $-$11.9 & \textbf{$-$4.4} \\
EmoFilm-EN    & 63.0 & 48.5 & $+$3.1  & $+$4.2 & \textbf{$+$4.8} \\
EMOVO         & 60.5 & 49.0 & $-$3.9  & $-$0.4 & \textbf{$+$3.0} \\
IEMOCAP       & 66.4 & 56.4 & $+$4.9  & $+$5.3 & \textbf{$+$5.4} \\
MSP-PODCAST   & 37.4 & 33.6 & $+$1.8 & $+$2.8 & \textbf{$+$3.3} \\
\hline
\emph{Average std} & & & 1.1 & 1.1 & 0.7 \\
\emph{Average} & & &
$-$1.2 & $+$2.0 & \textbf{$+$4.0} \\
\hline
\end{tabular}}
\end{table}
\begin{table}[t!]
\centering
\caption{Change in target speaker-level accuracy (\%) for the noisy adaptations on PC-GITA using DANN, CDAN, and the proposed approach. The accuracy of the baseline unadapted model on the source test set (Source) and on the target test set (Target) is also presented.}
\label{tab:pcgita_noisy}
\setlength{\tabcolsep}{10.5pt}
\scalebox{0.8}{\begin{tabular}{lccccc}
\hline
Corpus & Source & Target & DANN & CDAN & Proposed \\
\hline
PC-GITA & 86.3 & 62.1 & $+$10.2 & $+$10.7 & \textbf{$+$14.6} \\
\emph{Std} & & & 1.6 & 1.1 & 0.7 \\
\hline
\end{tabular}}
\end{table}
\textbf{Noisy adaptation for SER and pathological speech detection.} \enspace
Table~\ref{tab:ser_noise} presents the performance for the noisy adaptation setting averaged across 20 noisy target domains for the seven SER source datasets, whereas Table~\ref{tab:pcgita_noisy} presents the corresponding results for PC-GITA. For SER, DALN outperforms DANN and CDAN on most datasets, with the exception of CREMA-D where DANN performs best. Overall, DALN achieves the largest average performance improvement across all seven datasets ($4.0$\%), outperforming CDAN ($2.0$\%) and DANN ($-1.2$\%).
However, since the improvements vary considerably across datasets, the paired $t$-test over the seven datasets does not indicate statistical significance at this sample size, neither against DANN ($t(6)=1.75$, $p=0.131$) nor against CDAN ($t(6)=1.70$, $p=0.141$).
On PC-GITA, DALN likewise outperforms both adaptation approaches, improving over the unadapted model by $14.6$\%, compared with $10.7$\% for CDAN and $10.2$\% for DANN. Treating the 20 noisy target domains as paired observations, DALN's advantage is statistically significant over both DANN ($t(19)=2.89$, $p=0.009$) and CDAN ($t(19)=3.95$, $p<0.001$).

In summary, these extensive experimental results show that DALN consistently achieves the largest improvement among the three adversarial methods, with a statistically significant advantage over DANN in three of the four settings and over CDAN in two. Although DALN is not uniformly the best method for every individual adaptation, it is the most frequently best-performing and most consistent method overall.

\section{Conclusion}
We proposed to adapt the discriminator-free adversarial learning network DALN to unsupervised domain adaptation for speech tasks, reusing the task classifier as an implicit discriminator through a nuclear-norm Wasserstein discrepancy. We evaluated this approach across intra-lingual, cross-lingual, and noisy adaptation settings for SER and pathological speech detection. Across all settings, DALN consistently achieved the highest average performance among the evaluated adversarial methods. These results support discriminator-free adversarial adaptation as a robust alternative to discriminator-based methods for speech classification under domain shift.

\section{Acknowledgments}
This work was supported by the Swiss National Science Foundation project CRSII5\_202228 on ``Characterisation of motor speech disorders and processes''.
Authors thank Valmir Syla for his contribution to the initial exploratory experiments that helped motivate this work.

\bibliographystyle{IEEEtran}
\bibliography{refs}

\end{document}